\documentclass[12pt]{article}
\usepackage[utf8]{inputenc}
\usepackage{amssymb}
\usepackage{amsthm}
\usepackage{amsmath}
\usepackage{booktabs}
\usepackage{fancyhdr}
\usepackage{graphicx}
\usepackage{natbib}
\usepackage{breqn}
\usepackage{dsfont}
\usepackage{bbold}
\usepackage{subcaption}
\usepackage[margin=1.25in]{geometry}
\usepackage{setspace}
\newtheorem{theorem}{Theorem}[section]

\theoremstyle{definition}
\newtheorem{definition}{Definition}[section]
\newtheorem{assumption}{Assumption}[section]
\newtheorem{remark}{Remark}[section]

\title{Estimation of Independent Component Analysis Systems}
\author{Vincent Starck\thanks{Financial support from the European Research Council (Starting Grant No. 852332) is gratefully acknowledged. I thank Susanne Schennach for many helpful comments. I am also grateful to participants at Brown university's seminar for constructive discussions while presenting an early version of the paper. 
}\thanks{%
V.Starck@lmu.de} \\ LMU München}
\date{\today}

\newcommand\equaldef{\mathrel{\overset{\makebox[0pt]{\mbox{\normalfont\tiny\sffamily def}}}{=}}}
 
\DeclareMathOperator*{\argmin}{arg\,min} 
\DeclareMathOperator{\vect}{vec}

\newcommand{\sumjn}{\sum_{j=1}^n}
\newcommand{\sumtT}{\sum_{t=1}^T}
\newcommand{\meantT}{\frac{1}{T} \sum_{t=1}^T}

\newcommand{\meansT}{\frac{1}{T} \sum_{s=1}^T}

\begin{document}

\maketitle\thispagestyle{fancy}

\begin{abstract}

Although approaches to Independent Component Analysis (ICA) based on characteristic function seem theoretically elegant, they may suffer from implementational challenges because of numerical integration steps or selection of tuning parameters. 
Extending previously considered objective functions and leveraging results from the continuum Generalized Method of Moments of Carrasco and Florens (2000), I derive an optimal estimator that can take a tractable form and thus bypass these concerns. The method shares advantages with characteristic function approaches -- it does not require the existence of higher-order moments or parametric restrictions -- while retaining computational feasibility and asymptotic efficiency. The results are adapted to handle a possible first step that delivers estimated sensors. Finally, a by-product of the approach is a specification test that is valuable in many ICA applications. The method's effectiveness is illustrated through simulations, where the estimator outperforms efficient GMM, JADE, or FastICA, and an application to the estimation of Structural Vector Autoregressions (SVAR), a workhorse of the macroeconometric time series literature.

\end{abstract}

\section{Introduction}

Independent Component Analysis (ICA; \citet{comon1994independent, eriksson2003identifiability}) is a popular method which finds applications in fields as diverse as signal processing, machine learning, or Structural Vector Autoregressions (SVAR). \medskip

The standard model posits that an observed vector (of ``sensors'') at time $t$, $\eta_t$, is generated through $\eta_t = \Theta \varepsilon_t$ where $\varepsilon_t$ has independent entries and $\Theta$ is an unknown $n \times n$ matrix. $\varepsilon_t$, sometimes referred as the ``sources'', is a vector that contains the latent factors that affect the system through the mixing matrix, $\Theta$. \medskip

Various methods have been proposed to uncover the unmixing matrix, $\Theta^{-1}$. Early methods typically attempted to maximize a measure of non-normality or use maximum entropy, often making use of third- and fourth-order moments. A popular and fast method based on non-Gaussianity is the fastICA algorithm \citep{oja2006fastica}. A list of algorithms and applications can be found in \citet{hyvarinen2001independent, agrawal2022blind}. \medskip

A broad estimation strategy is to rely on maximum likelihood or related methods. Many papers \citep{bach2002kernel, samarov2004nonparametric, chen2006efficient, ilmonen2011semiparametrically, samworth2012independent, ablin2018faster} have assumed parametric, smooth, or log-concave densities to devise an estimation strategy for the unmixing matrix. Nevertheless, misspecification bias is a concern as the family of distributions is typically unknown, and assumptions of smoothness, unimodality, or absence of atoms are not innocuous in applications. In addition, many of these approaches require a choice of tuning parameter or are not straightforward to implement. Efficient approaches to ICA remain relatively rare, although a few have been proposed (\textit{e.g.}, \citet{chen2006efficient}). \medskip 

I propose a nonparametric approach to estimate the unmixing matrix based on the empirical characteristic function that does not require the existence of higher-order moments or distributional restrictions and can achieve asymptotic efficiency. I also explicitly allow $\eta_t$ to be (consistently) estimated rather than directly observed to account for vanishing noise or an estimation step, as happens, for instance, in Structural Vector Autoregressions (SVAR) applications where the ICA system requires inputs from a first-step regression. \medskip

\citet{eriksson2003characteristic} (see also \citet{chen2005consistent} for asymptotic properties) previously obtained an identifying equation function in terms of the characteristic function and proposed an integration scheme to estimate ICA systems. This paper shows that refining their objective function enables the derivation of an optimal weighting scheme, leading to an efficient estimator that can bypass numerical integration steps, is compatible with noise due a first step such as autoregressions in SVAR, and delivers a test of the ICA system's validity. \medskip

\section{Estimation of ICA systems}
\subsection{Identification of $\Theta$}

Identification of systems of linear combinations of unobserved independent variables has been extensively discussed in the literature and extended to more general systems \citep{reiersol1950identifiability, comon1994independent, bonhomme2009consistent, ben2016identification}. The ICA system remains a workhorse theoretical model and is routinely used in empirical applications. A general result on the identification of such systems is provided by \citet{eriksson2003identifiability}.

%If an observable vector X is generated through X = AS where A is a $p \times m$ constant matrix and $S$ is a vector of independent real-valued variable, then $A$ is identified if (i) no variable is $S$ is normally distributed or (ii) $A$ is of full column rank and at most one variable in $S$ is normally distributed. \medskip

%\citet{eriksson2003identifiability} also establish that condition (ii) is sufficient to identify the distribution of $S$. In the 'square' ICA framework, $\Theta$ is a full rank-matrix so that, provided at most one source is normal, identification is achieved. \medskip 

In the square matrix case, identification is easily established by noting that two observationally equivalent systems $(\Theta, \varepsilon)$ and $(\Theta^{*}, \varepsilon^{*})$ must satisfy $\Theta  \varepsilon \overset{d}{=} \Theta^{*} \varepsilon^{*}$ and thus $\varepsilon \overset{d}{=} \Theta^{-1} \Theta^{*} \varepsilon^{*}$, where $\overset{d}{=}$ denotes equality in distribution. But if $\varepsilon$ is to have independent entries, the Darmois–Skitovich theorem \citep{darmois1953analyse, skitovitch1953property} requires trivial linear combinations in the absence of normality. As a result, $\Theta^{-1} \Theta^{*}$ must be a permutation matrix (possibly scaled). This well-known result is summarized in the following theorem:

\begin{theorem}[Identification]\label{theorem_identification}
Consider the system $\eta_t = \Theta \varepsilon_t$.
$\Theta$ is identified up to column scale and permutations from the distribution of $\eta_t$ if (i) it is invertible, and (ii) the vector $\varepsilon_t$ contains independent random variables among which at most one is normal.
\end{theorem}

Since identification is obtained only up to scale and column permutations, a normalization and an order are still needed. In the absence of application-specific knowledge to assign identities to shocks, the choice can be made out of convenience. \medskip

I use a unit norm normalization for each column of $\Theta$ and denote the corresponding compact parameter space by $\overline{\Theta}$. The constraint can be explicited through polar coordinates. For instance, in the $2 \times 2$ case, $\Theta = \begin{pmatrix} \cos(\theta_1) & \cos(\theta_2) \\ \sin(\theta_1) & \sin(\theta_2) \end{pmatrix}$, where $\theta_1$ and $\theta_2$ lie between 0 and $\pi$. Columns can be easily ordered by setting, say, $\theta_1 < \theta_2$, with straightforward adaptations to higher-dimensional settings using lexicographic ordering. As a result, the properties can equivalently be discussed in terms of $\theta$, under the bijective mapping $\Theta = \Theta(\theta)$. \medskip

In what follows, it is assumed that the system is identified by imposing conditions (i) and (ii) of Theorem \ref{theorem_identification} and the normalizations described above.

\subsection{The class of estimators based on characteristic functions}

I will make use of the following notation. I define $\vec{s}$ to be a $1 \times n$ row vector. $\varphi_X$ denotes the characteristic function of the random vector $X$, \textit{i.e.} $\varphi_X(\vec{s}) \equaldef \mathds{E}[e^{i\vec{s}x}]$. $\varphi$ (without a subscript) refers to the characteristic function of $\eta$. Real and imaginary parts are denoted by $\Re$ and $\Im$, respectively. The $j^{th}$ column of $\Theta$ is an $n \times 1$ vector denoted by $\Theta_{\cdotp j}$. 

%; the $j^{th}$ row of $\Theta$ is a $1 \times n$ vector denoted by $\Theta_{j \cdotp}$.

\subsubsection{The estimator}

By independence of sources, the observed variables' distribution is related to the distribution of their unobserved counterparts through

\begin{equation}
    \varphi(\vec{s}) = \prod_{j=1}^n \varphi_{\varepsilon_{j}}(\vec{s} \Theta_{\cdotp j})
\end{equation}
while each source's characteristic function can be recovered from that of the sensors via

\begin{equation}
    \varphi_{\varepsilon_{j}}(s) = \varphi(s \Theta_{j \cdotp}^{-1})
\end{equation}
where $\Theta_{j \cdotp}^{-1}$ is the $j$-th row in $\Theta^{-1}$. \medskip 

A functional equation for the characteristic function of $\eta$ in terms of the unknown $\Theta$ can be obtained using the last two expressions. First, define $P_j \equaldef \Theta_{\cdotp j} \Theta_{j \cdotp}^{-1}$ whose immediate properties are $P_j P_k = \mathds{1}_{j=k} P_j \ \forall j, k$, $\sum_{j=1}^n P_j = I_n = \sum_{j=1}^n P_j^{'}$, and $\mbox{rank}(P_j) = 1$. \medskip

In addition, the collection of $P_j$ is isomorphic to $\Theta$ once a normalization on $\Theta$ is imposed. Next, substituting (2) into (1) yields an expression which directly links the characteristic function of $\eta$ to $\Theta$:

\begin{equation}\label{eq:1}
    \varphi(\vec{s}) = \prod_{j=1}^n \varphi(\vec{s} P_j)
\end{equation}
This result was also obtained by \citet{eriksson2003characteristic}. The identity can be expressed as the condition $q(\vec{s}, \theta) \equaldef \varphi(\vec{s}) - \prod_{j=1}^n \varphi(\vec{s} P_j) = 0$ for all $\vec{s}$. \medskip
\begin{remark}
    If one assumes that the characteristic function does not vanish\footnote{Assumptions excluding zeros are common, \textit{e.g.}, literature on nonparametric deconvolution (see \citet{schennach2004estimation} and references therein).} or restricts the analysis to a neighborhood of the origin, then the criterion can alternatively be expressed in terms of the cumulant generating function: $\sum_{j=0}^n a_j \ln\left(\frac{1}{T} \sum_{t=1}^T e^{i \vec{s} P_j \eta_t}\right) = 0$ with $a_j = (-1)^{\mathds{1}_{j > 0}}$ and $P_0 = I$. Both criteria can be used to form an optimally weighted estimator of $\Theta$ with closely related expressions; the log form, however, requires more care to handle zeros.
\end{remark} 

Let $\hat{\varphi}(\vec{s}) \equaldef \meantT e^{i \vec{s} \eta_t}$ be the empirical counterpart of $\varphi_\eta$ and let $q_T(\vec{s}, \theta) \equaldef \hat{\varphi}(\vec{s}) - \prod_{j=1}^n \hat{\varphi}(\vec{s} P_j)$. \citet{eriksson2003characteristic} consider a criterion based on minimizing integrals of the form  
\begin{equation}
    \int \vert q_T(\vec{s}, \theta)\vert^2 w(\vec{s}) \mbox{d}\vec{s} = \int (\Re q_T(\vec{s}, \theta)^2 + \Im q_T(\vec{s}, \theta)^2) w(\vec{s}) \mbox{d}\vec{s}
\end{equation}
using some weight function $w$. They then propose tractable weighting schemes to facilitate integration. The approach is neat, but suffers from a couple of shortcomings. First, it does not allow for weighting interactions at different points $\vec{s}$, which precludes efficiency because neighboring points contain similar information. Second, the weights are chosen suboptimally in order to avoid numerical integration, which can lead to further efficiency loss. \medskip

In analogy with the formation of a quadratic form for estimating equations or Generalized Method of Moments (GMM; \citet{hansen1982large}), consider 
\begin{align}\label{objective}
    Q_T(\theta) = \int \int \begin{pmatrix} \Re q_T(\vec{s}, \theta) & \Im q_T(\vec{s}, \theta)\end{pmatrix} W(\vec{s}, \vec{r}) \begin{pmatrix} \Re q_T(\vec{r}, \theta) \\ \Im q_T(\vec{r}, \theta)\end{pmatrix} \pi(\mbox{d}\vec{r}) \pi(\mbox{d}\vec{s})
\end{align}
where $\pi$ is a probability measure and $W$ is a (symmetric, positive semi-definite) weighting matrix. \medskip

This criterion can be induced by a linear operator $B$ by considering the norm $\Vert B q_T\Vert_H$, where $H$ is a Hilbert space of square-integrable functions with the scalar product $<f, g> = \int f(\vec{s}) \overline{g(\vec{s})} \pi(\vec{s})$. This is similar to \citet{carrasco2000generalization, carrasco2002efficient}'s objective function for a continuum of moment conditions.\footnote{A difference is that $q_T$ is obtained from a transformation of moments. Moment conditions can be obtained by introducing characteristic functions of the shocks as nuisance parameters. Up to negligible terms, the solution uses empirical characteristic functions and thus coincides with the nonlinear system.} The following section establishes the asymptotic properties of estimators obtained by minimizing (\ref{objective}) for any weighting matrix. The weighting matrix $W_T$ is allowed to be an estimate of some target matrix $W$. The results are also made applicable to the case where $\eta_t$ is not observed but can be obtained from a first step, as in SVAR applications.

%Here, the choice of $\pi$ is mostly inconsequential -- asymptotically, any $\pi > 0$ will lead to equivalent estimators -- while the choice of $W$ possibly determines optimality.

\subsubsection{Asymptotics}

%$|| B g || \leq \sup_{s, t} W \int \int g_s \overline{g}_t = C \vert \int g \vert^2 \leq  \int \vert g \vert^2$

To establish consistency, it is necessary to exclude sequences toward degenerate matrices if $\eta_t$ is to be estimated. I make use of the following assumption, which slightly strengthens the invertibility assumption by bounding the matrix an $\epsilon$ away from degeneracy. \medskip

\begin{definition}[$\epsilon$-invertibility]
The angle between any two columns of $\Theta$ is of at least $\epsilon > 0$.
\end{definition}

In some applications, the $\eta_t$'s are directly observed, and consistency can be established under the usual full-rank condition on $\Theta$. \medskip

In what follows, $\theta_0$ refers to the true value of $\theta$ and I let $q_0(\vec{s}) \equaldef q(\vec{s}, \theta_0)$. The proposed estimator is consistent for the true value $\theta_0$, as summarized by the following Theorem, which is proven in the Appendix. \medskip

\begin{theorem}[Consistency] 
The estimator $\hat{\theta} \equaldef \argmin_{\theta} Q_T(\theta)$ is consistent for $\theta_0$ if (i) there is a consistent estimator, $\hat{\eta}_t$, of $\eta_t$ that satisfies $\meantT \vert \hat{\eta}_t - \eta_t\vert \rightarrow^p 0$ and $\Theta = \Theta(\theta)$ is $\epsilon$-invertible, and (ii) $W_T$ converges uniformly to a positive definite matrix $W$ that satisfies $\sup_{\vec{s}, \vec{r}} W_{ij}(\vec{s}, \vec{r}) < \infty$.
\end{theorem}

I now turn to the derivation of the asymptotic distribution of the estimator. 
\begin{theorem}[Asymptotic Normality]\label{asymptotic_normality}
$\hat{\theta}$ is asymptotically normally distributed. 
Specifically, if (i) $\hat{\eta}_t$ is obtained from a first-stage where $\eta_t$ is an independent error term so that $\hat{\eta_t} - \eta_t = w_t' (\hat{\beta} - \beta)$ for some consistent estimator $\hat{\beta}$ of $\beta$ and $\Theta = \Theta(\theta)$ is $\epsilon$-invertible, (ii) $W_T$ converges uniformly to a positive definite matrix $W$ that satisfies $\sup_{\vec{s}, \vec{r}} W_{ij}(\vec{s}, \vec{r}) < \infty$, and (iii) $\varepsilon_t$ has second moments.

Then, $\sqrt{T} (\hat{\theta} - \theta_0) \overset{d}{\rightarrow} N(0; B V B')$ where 

\begin{equation}
    B \equaldef \left[\int  \begin{pmatrix} \Re\frac{\partial q_0}{\partial \theta} (\vec{r}) & \Im(\frac{\partial q_0}{\partial \theta} (\vec{r}) \end{pmatrix} W(\vec{r}, \vec{s}) \begin{pmatrix} (\Re \frac{\partial q_0}{\partial \theta} (\vec{s}))^' \\ (\Im \frac{\partial q_0}{\partial \theta} (\vec{s}))^' \end{pmatrix} \pi(\mbox{d}\vec{r}) \ \pi(\mbox{d}\vec{s}) \right]^{-1}
\end{equation}
and
\begin{align}
\begin{split}
    V \equaldef \int \int & \begin{pmatrix}\Re\frac{\partial q_0}{\partial \theta} (\vec{r}) & \Im\frac{\partial q_0}{\partial \theta} (\vec{r})\end{pmatrix} W(\vec{r}, \vec{s}) K(\vec{r}, \vec{s}) \\
    & W(\vec{r}, \vec{s})' \begin{pmatrix}(\Re\frac{\partial q_0}{\partial \theta} (\vec{s}))' \\ (\Im\frac{\partial q_0}{\partial \theta} (\vec{s}))'\end{pmatrix} \ \pi(\mbox{d}\vec{r}) \ \pi(\mbox{d}\vec{s})
\end{split}
\end{align}

\noindent where $K(\vec{r}, \vec{s})$ is the covariance function for the real and imaginary parts of $q_0$. 
\end{theorem}

As usual, the iid assumption can be weakened to ergodicity and strict stationarity. This is done with a natural adaptation of the proof, noting that results about convergence of characteristic functions have generalizations to ergodic, stationary settings \citep{feuerverger1990efficiency}. Two corrections may apply to the asymptotic variance. First, the use of a long-run variance might be warranted since uncorrelatedness of sensors does not translate to that of their empirical characteristic functions. Second, estimation of $\eta_t$ must be accounted for since its disappearance hinges on the vanishing of the term $\frac{\partial q_0(\eta_t(\overline{\beta}))}{\partial \beta} (\hat{\beta} - \beta)$, which relies on independence. \medskip

Finally, since $\Theta$ is usually the parametrization of interest, the Delta method can be applied and asymptotic normality ensues for the corresponding estimator with the same asymptotic variance scaled by $\partial_\theta \Theta(\theta)$. The asymptotic variance is analyzed in details in Appendix B.

\section{Efficient estimation}

\subsection{Optimal objective function}

    %Q_T^E(\theta) = \int  \int  \begin{pmatrix} \Re q_T(\vec{s}, \theta) & \Im q_T(\vec{s}, \theta)\end{pmatrix} W_T(\vec{s}, \vec{r}) \begin{pmatrix} \Re q_T(\vec{r}, \theta) \\ \Im q_T(\vec{r}, \theta)\end{pmatrix} \mbox{d}\vec{r} \mbox{d}\vec{s}
%Now $W(\vec{s}, \vec{r})$ possesses the necessary additional degree of freedom to hope for cancellation between $A$ and $V$.

Since the asymptotic variance in Theorem \ref{asymptotic_normality} has the common ``sandwich'' form, a natural choice of weighting function consists in inverting the covariance term in order to minimize the asymptotic variance. This is reminiscent of the choice of efficient matrix in GMM \citep{hansen1982large} and its extension to a continuum of moments conditions developed by \citet{carrasco2000generalization}. \medskip

As \citet{carrasco2000generalization} establish, efficient estimation in the continuous case requires inverting the covariance operator $C: h \rightarrow \int K(\vec{r}, \vec{s}) h(\vec{s}) \mbox{d}\vec{s}$ which is not possible on the whole reference space. They propose a regularized sample version $C_T^{\alpha_T}$ where $\alpha_T$ is a smoothing parameter that disturbs the eigenvalues of $C$. The choice of $\alpha_T$ has been discussed in subsequent papers, see \citet{carrasco2017efficient} and \citet{amengual2020testing}. \medskip

%While $\Re q_T(\vec{s}, \theta)$ are $\Im q_T(\vec{s}, \theta)$ are not \textit{per se} sample moments, they do have a zero asymptotic counterpart and the results of \citet{carrasco2000generalization} can be adapted to the present framework. \medskip 

%Specifically, denoting real and imaginary parts of $q_T$ by $g_j, j=1,2$, their objective function $\Vert B_n(q_T(\theta))\Vert$ matches $Q_T$ when $B_n$ is an integral operator $(B_n g) (\vec{s}) = \left(\sum_{l=1}^2 \int  b^{jl}(\vec{s}, \vec{r}) g_{l}(\vec{r}) \mbox{d}\vec{r}\right)_{j=1, 2}$ that generates a weighting matrix through $W_T^{jk}(\vec{s}, \vec{r}) = \sum_{l, l'=1, 2} \int  b^{jl}(\vec{u}, \vec{s}) b^{jl}(\vec{u}, \vec{r}) d\vec{u}$. \medskip 

The optimal estimator minimizes 

\begin{equation}\label{efficient_objective}
    Q_T^* \equaldef \sum_{t=1}^T \frac{\mu_{t;T}}{\mu_{t;T}^2 + \alpha_T} \left\vert<q_T, \chi_{t;T}>\right\vert^2
\end{equation}
where $\lambda_{t;T}$ and $\chi_{t;T}$ are the eigenvalues and eigenfunctions of $C_T$, the sample counterpart to $C$. These can be easily deduced from the eigen-decomposition of a matrix $M$, as explained in the next subsection. \medskip 

Moreover, under the assumptions of Theorem 3.2 and provided $\alpha_T \rightarrow 0$ while $\alpha_T^3 T \rightarrow \infty$, the expected simplification of the asymptotic variance occurs so that the asymptotic distribution becomes

\begin{equation}
    \sqrt{T}(\hat{\theta} - \theta_0) \overset{d}{\rightarrow} N\left(0; \left|\left|\frac{\partial q_0}{\partial \theta}\right|\right|_{C}^{-2}\right) \overset{d}{=} N\left(0; B\right)
\end{equation}
where the weighting matrix used to compute $B$ is now based on the inverted covariance operator. \medskip 

Although the objective function appears to induce a computationally intensive procedure due to multiple integrations, in practice the estimator is obtained by minimizing equation ($\ref{efficient_objective}$) and the main computational burden arises from evaluating a matrix and computing its eigen-decomposition.
Furthermore, using the efficient form of the estimator carries significant benefits. It removes the need to specify an arbitrary form of the weighting matrix and furthers efficiency. In particular, \citet{carrasco2000generalization} show that the use of a continuum of moment conditions closes the efficiency gap between GMM and MLE. \medskip

%Fully exploiting the distributional information implies using the full extent of zero conditions, \textit{i.e.} asymptotically integrating over the whole space. This is backed up by the simplified form of the asymptotic variance, which decreases as $\Omega$ expands. As optimal weighting necessarily cancels infinite variances brought by the zeros of $f$, arbitrarily expanding the integration region becomes possible, though \citet{carrasco2000generalization}'s assumption that $\mathds{E}[\Vert f\Vert^4] < \infty$, required to derive the asymptotic distribution of the efficient estimator, is no longer a trivial assumption in presence of zeros and unbounded space. \medskip

The efficiency follows from results in \citep{carrasco2002efficient} by completeness of characteristic functions, once a few regularity conditions hold:

\begin{assumption}[Sufficient conditions for efficiency]
    Each shock admits a continuously-differentiable density and has finite variance.
\end{assumption}
% Density of observables is a convolution of scaled shock densities (theta f(x/theta)). Need \int \sup_theta \vert \partial f/partial \theta dx to be finite. Theta lives in a compact, if the decay is sufficiently fast in the tails to ensure second moments and the derivative is continuous, this holds.
Simulations show a good behavior of the method in general, although guarantees of optimality follow from the theory only in the continuous case.

\subsection{Implementation}

Let $\Psi^\tau$ denote the moment conditions at observation $\tau$. The covariance operator reads
\begin{align*}
    \begin{split}
    (C g)(\vec{r}) & = \int \frac{1}{T} \sum_{\tau=1}^T \Psi^\tau(\vec{r}) \overline{\Psi^\tau(\vec{s})}' g(\vec{s}) \ \pi(\mbox{d}\vec{s}) \\
    & = \frac{1}{T} \sum_{\tau=1}^T \Psi^\tau(\vec{r}) \int \overline{\Psi^\tau(\vec{s})}' g(\vec{s}) \ \pi(\mbox{d}\vec{s})
    \end{split}
\end{align*}
which implies that the eigenfunctions take the form 
\begin{equation}
    \chi(\vec{r}) = \frac{1}{T} \sum_{\tau=1}^T c_{\tau}  \Psi^\tau(\vec{r})
\end{equation}

The eigenvalues and coefficients are given by the eigenvalue-eigenvector pairs of $M/T$, where the elements of $M$ are given by

\begin{equation}
    M_{\tilde{\tau} \tau} = \int \overline{\Psi^\tau(\vec{s})}'  \Psi^{\tilde{\tau}}(\vec{s}) \ \pi(\mbox{d}\vec{s})
\end{equation}

The eigenvector/eigenvalue couples $(c, \lambda)$ of $M/T$ are then used to form orthonormalized\footnote{The eigenfunctions are automatically orthogonal, but need to be normed.} eigenfunctions $\chi_{t, T}, t=1, \ldots, T$. 
%and $B$ is the basis $\{\sum_{k=0}^n a_k \phi_k(\vec{r}) e^{i \vec{r} P_k \eta_{\tau}}\}$. 

%Then $<f, \hat{\chi}> = E' <f, B>$. 

%If $\iota' A  = 0$, then $\lambda_i = 0$ or $sum(e_i) = 0$. If $A$ is symmetric, it is also true for $A \iota = 0$.
%This is because $\iota' A x = \sum \lambda_i (\iota' e_i) (v_i' x) = 0$ where $v_i$ is from the inverted eigenvector matrix. This is true for any $x$, so one can get $\lambda_i (\iota' e_i) = 0$ for any $i$ by taking $x$ as eigenvector.

%Mx=lambda x; if sum(x) is 0 then (M x - M iota iota') x = lambda x too so the eigenvectors are common. This leads to M10 having the same eigenvectors as M00 as long as eigenvalues are nonzero. 

%Relationships between $r_{r_k}$ and $r_{t_k}-1$ linear combinations: a linear combination of the latter with coefficients $\vec{c}$ is equivalent to using $\vec{c} - \vec{c} \iota$ since the $r$ sum up to 1.   
%the sample version of $\int \sum_{j=0}^n \sum_{k=0}^n a_j a_k \left(\frac{e^{i \vec{s}_j \eta_\tau}}{\meantT e^{i \vec{s} P_j \eta_t}} - 1\right) \overline{\sum_{k=0}^n a_k \left(\frac{e^{i \vec{s}_k \eta_{\tilde{\tau}}}}{\meantT e^{i \vec{s} P_k \eta_t}}\right)} \vert \phi(\vec{s}) \vert^2 \ \pi(\mbox{d}\vec{s})$ reads
\bigskip

Computing (\ref{efficient_objective}) thus requires two main objects: the matrix $M$ and the scalar products $<q_T, \chi_{t;T}>$. Their computation can be significantly simplified because the choice of integration measure $\pi$ does not affect asymptotic properties \citep{carrasco2017efficient}. Hence, as long as $\pi > 0$ over the whole space, ensuring tractability for ease of implementation is the main concern. I propose to integrate with respect to a Gaussian, \textit{i.e.}, $\pi(\mbox{d}\vec{s}) = (2 \pi)^{-n/2} e^{-\frac{1}{2} \Vert \vec{s} \Vert^2}$. Then, one can compute $M$ whose entries read
\begin{align*}
    \begin{split}
        \int & \overline{\Psi^\tau(\vec{s})}'  \Psi^{\tilde{\tau}}(\vec{s}) \ \pi(\mbox{d}\vec{s}) \\ 
        = & \mbox{exp}(-0.5 \ \left\Vert \eta_{\tau} - \eta_{\tilde{\tau}}\right\Vert^2) + \sum_{j=1}^n \mbox{exp}(-0.5 \ \left\Vert P_j(\eta_{\tilde{\tau}} - \eta_{\tau})\right\Vert^2) \\
        - & 2 \frac{1}{T^n} \sum_{\{\tilde{\tau}_k\}} \mbox{exp}\left(-0.5 \ \left\Vert \sum_{k=1}^n P_k \eta_{\tilde{\tau}_k} - \eta_{\tau}\right\Vert^2\right) - 2 \frac{1}{T} \sum_{\tilde{\tau} = 1}^T \sumjn \mbox{exp}\left(-0.5 \ \left\Vert P_j (\eta_{\tilde{\tau}} - \eta_{\tau}) \right\Vert^2\right) \\
        + & \frac{1}{T^{2n}} \sum_{\{\tau_j, \tilde{\tau}_k\}} \mbox{exp}\left(-0.5 \ \left\Vert \sum_{k=1}^n P_k \eta_{\tilde{\tau}_k} - \sumjn P_j \eta_{\tau_j}\right\Vert^2\right) + \frac{1}{T^2} \sum_{\tau=1}^T \sum_{\tilde{\tau}=1}^T \sumjn \mbox{exp}\left(-0.5 \left\Vert P_j (\eta_{\tilde{\tau}} - \eta_\tau)\right\Vert^2\right)
    \end{split}
\end{align*}
noting that $\int e^{\vec{s} (a+b i)} \pi(\mbox{d}\vec{s}) = e^{\Vert a \Vert^2/2} e^{a' b i} e^{- \Vert b \Vert^2/2}$.

Similarly, one can obtain the scalar products as linear combinations of 
$<\Psi, \Psi^\tau > = \meantT \int \overline{\Psi^\tau(\vec{s})}'  \Psi^{t}(\vec{s}) \pi(\mbox{d}\vec{s})$. \medskip

Therefore, computing the objective function does not require numerical integration. The averages over all pairs of time can be high-dimensional when $T$ and $n$ get large, so that drawing times at random and computing the resulting sample average may be desirable to lower the computational cost.

%& = \sum_{j,k \neq 0} \frac{1}{T^{2(n-1)}} \sum_{\{t_j, s_k\}} \mbox{exp}(-0.5 \ \Vert P_j \eta_{\tau} - P_k \eta_{\tilde{\tau}} + \sum P_{\tilde{j}} \eta_{t_{\tilde{j}}} - \sum P_{\tilde{k}} \eta_{s_{\tilde{k}}} \Vert^2) \\ 
       % & - \sum_{k \neq 0} \frac{1}{T^{n-1}} \sum_{\{s_{\tilde{k}\}}} \mbox{exp}(-0.5 \ \Vert \eta_\tau - P_k \eta_{\tilde{\tau}} - \sum P_{\tilde{k}} \eta_{s_{\tilde{k}}} \Vert^2) \\
        %& - \sum_{j \neq 0} \frac{1}{T^{n-1}} \sum_{\{t_{\tilde{j}}\}} \mbox{exp}(-0.5 \ \Vert P_j \eta_\tau - \eta_{\tilde{\tau}} - \sum P_{\tilde{k}} \eta_{t_{\tilde{k}}} \Vert^2) \\

\subsection{Tests}

Asymptotic normality provides the basis for usual confidence intervals and tests. Moreover, an advantage of the analogy with GMM is the potential for a specification test, in the spirit of over-identifying restrictions. As detailed in \citet{carrasco2000generalization}, such a test can be constructed on the basis of 

\begin{equation}
    \frac{||\sqrt{T} Q_T^*||_{C_T^{\alpha_T}}^2 - \sum_{t=1}^T\frac{\lambda_{t;T}^2}{\lambda_{t;T}^2 + \alpha_T}}{\sqrt{2 \sum_{t=1}^T\frac{\lambda_{t;T}^4}{(\lambda_{t;T}^2 + \alpha_T)^2}}} \overset{d}{\rightarrow} N(0;1)
\end{equation}

provided $\alpha_T \sum_{t=1}^T \frac{\lambda_{t;T}^4}{(\lambda_{t;T}^2 + \alpha_T)^2} \rightarrow \infty$ and the assumptions for asymptotic normality of the efficient estimator hold. \medskip

Such a test is valuable when working with ICA systems as it provides a feedback about the validity of the entire structure.

\section{Simulations}

I generate samples of $\eta_t$ through equation $\eta_t = \Theta \epsilon_t$ with various distributions for the epsilons and a sample size of $T=150$. I compare the performance in recovering the lag polynomial of the efficient estimator of Section 4 to that of efficient GMM based on moment conditions (\textit{i.e.} deriving identifying equations implied by independence under the assumptions that moments up to order 4 exists, see \textit{e.g.}, \citet{guay2018identification}) and fastICA \citep{oja2006fastica}. In the forthcoming tables, the corresponding estimators are denoted by log-cf, GMM, and fICA, respectively. \medskip

I consider the following distributions for the sources: student with 3 degrees of freedom, uniform on [-1;1], Binomial(20, 0.3), and Gamma(5, 1/7). All distributions are centered as to have mean zero. These distributions account for a variety of cases such as fat tails, skewness, or presence of atoms.  \medskip 

Consider first a student distribution with 3 degrees of freedom. In this case, the student distribution exhibits fat tails and moments higher than 2 do not exist, endangering identification strategies based on higher moments. \medskip

\begin{table}[!ht]\caption{Student distribution $\nu=3$}
\[\begin{array}{l | l | l | l | l | l | l | l | l | l}
\toprule
    {} & \multicolumn{3}{c}{\mbox{Bias}} & \multicolumn{3}{c}{\mbox{Standard deviation}} & \multicolumn{3}{c}{\mbox{RMSE}}\\ 
\midrule
	\Theta_{\cdot 1} & \mbox{log-cf} & \mbox{GMM} & \mbox{fICA} & \mbox{log-cf} & \mbox{GMM} & \mbox{fICA} & \mbox{log-cf} & \mbox{GMM} & \mbox{fICA}\\
	 0.71 & -0.05 & -0.02 & -0.31 & 0.17 & 0.22 & 0.55 & 0.17 & 0.22 & 0.63 \\ 
	 0.71 & 0.01 & -0.07 & 0.01 & 0.13 & 0.28 & 0.17 & 0.14 & 0.29 & 0.17 \\ 
    \Theta_{\cdot 2} & \mbox{log-cf} & \mbox{GMM} & \mbox{fICA} & \mbox{log-cf} & \mbox{GMM} & \mbox{fICA} & \mbox{log-cf} & \mbox{GMM} & \mbox{fICA}\\
     -0.50 & -0.03 & 0.03 & 0.11 & 0.17 & 0.28 & 0.55 & 0.18 & 0.28 & 0.56 \\
     0.87 & -0.04 & -0.05 & -0.15 & 0.10 & 0.19 & 0.17 & 0.11 & 0.20 & 0.22 \\
\bottomrule
\end{array}\]
\end{table}

\medskip

It appears the estimator based on log-empirical characteristic function considerably outperforms both efficient GMM and fastICA estimators when the sources are student distributed. The gains come mostly from a lower standard deviation, though there is some bias reduction especially compared to fastICA. \medskip 

Now, I turn to uniform and binomial distributions. Both distribution have all their moments but one is continuous and symmetric while the other is discrete and skewed. Both efficient GMM and the characteristic-function based estimator outperform fastICA for the uniform distribution. In the binomial case, the characteristic function based estimator again fares better than both efficient GMM and fastICA, with a considerable reduction in mean square error originating from lower standard deviations. \medskip 

\begin{table}[!h]\caption{Uniform distribution}
\[\begin{array}{l | l | l | l | l | l | l | l | l | l}
\toprule
    {} & \multicolumn{3}{c}{\mbox{Bias}} & \multicolumn{3}{c}{\mbox{Standard deviation}} & \multicolumn{3}{c}{\mbox{RMSE}}\\ 
\midrule
\Theta_{\cdot 1} & \mbox{log-cf} & \mbox{GMM} & \mbox{fICA} & \mbox{log-cf} & \mbox{GMM} & \mbox{fICA} & \mbox{log-cf} & \mbox{GMM} & \mbox{fICA}\\
	 0.71 & -0.04 & -0.10 & -0.28 & 0.14 & 0.16 & 0.57 & 0.15 & 0.19 & 0.63 \\ 
	 0.71 & 0.01 & 0.06 & -0.01 & 0.12 & 0.11 & 0.12 & 0.12 & 0.13 & 0.12 \\ 
    \Theta_{\cdot 2} & \mbox{log-cf} & \mbox{GMM} & \mbox{fICA} & \mbox{log-cf} & \mbox{GMM} & \mbox{fICA} & \mbox{log-cf} & \mbox{GMM} & \mbox{fICA}\\
     -0.50 & -0.01 & -0.06 & 0.10 & 0.15 & 0.20 & 0.58 & 0.15 & 0.21 & 0.59 \\
     0.87 & -0.02 & -0.07 & -0.17 & 0.08 & 0.11 & 0.12 & 0.08 & 0.13 & 0.21 \\
\bottomrule
\end{array}\]
\end{table}
\begin{table}[!h]\caption{Binomial distribution $n=20, p=0.3$}
\[\begin{array}{l | l | l | l | l | l | l | l | l | l}
\toprule
    {} & \multicolumn{3}{c}{\mbox{Bias}} & \multicolumn{3}{c}{\mbox{Standard deviation}} & \multicolumn{3}{c}{\mbox{RMSE}}\\ 
\midrule
	\Theta_{\cdot 1} & \mbox{log-cf} & \mbox{GMM} & \mbox{fICA} & \mbox{log-cf} & \mbox{GMM} & \mbox{fICA} & \mbox{log-cf} & \mbox{GMM} & \mbox{fICA}\\
	 0.71 & -0.02 & -0.06 & -0.30 & 0.13 & 0.27 & 0.51 & 0.13 & 0.28 & 0.60 \\
     0.71 & 0.01 & -0.07 & -0.01 & 0.14 & 0.30 & 0.31 & 0.14 & 0.31 & 0.31 \\
	 \Theta_{\cdot 2} & \mbox{log-cf} & \mbox{GMM} & \mbox{fICA} & \mbox{log-cf} & \mbox{GMM} & \mbox{fICA} & \mbox{log-cf} & \mbox{GMM} & \mbox{fICA}\\
	 -0.50 & -0.01 & 0.00 & 0.12 & 0.18 & 0.32 & 0.53 & 0.18 & 0.32 & 0.54 \\ 
     0.87 & -0.02 & -0.11 & -0.17 & 0.08 & 0.27 & 0.31 & 0.08 & 0.29 & 0.35 \\ 
\bottomrule
\end{array}\]
\end{table} \medskip

Finally, the last tables show more contrasted results. In the case of a gamma distribution, log-cf and fICA estimators exhibit similar performance in terms of RMSE and tend to be outperformed by efficient GMM. The characteristic function based estimator occasionally displays a greater bias, which reduces its performance with these distributions, at least for some parameters. \medskip

\begin{table}[!ht]\caption{Gamma distribution $\alpha =5, \beta=1/7$}
\[\begin{array}{l | l | l | l | l | l | l | l | l | l}
\toprule
    {} & \multicolumn{3}{c}{\mbox{Bias}} & \multicolumn{3}{c}{\mbox{Standard deviation}} & \multicolumn{3}{c}{\mbox{RMSE}}\\ 
\midrule
	\Theta_{\cdot 1} & \mbox{log-cf} & \mbox{GMM} & \mbox{fICA} & \mbox{log-cf} & \mbox{GMM} & \mbox{fICA} & \mbox{log-cf} & \mbox{GMM} & \mbox{fICA}\\
	 0.71 & -0.20 & -0.14 & -0.29 & 0.56 & 0.34 & 0.53 & 0.59 & 0.36 & 0.61 \\
     0.71 & -0.15 & -0.02 & -0.01 & 0.35 & 0.30 & 0.25 & 0.39 & 0.31 & 0.25 \\
	 \Theta_{\cdot 2} & \mbox{log-cf} & \mbox{GMM} & \mbox{fICA} & \mbox{log-cf} & \mbox{GMM} & \mbox{fICA} & \mbox{log-cf} & \mbox{GMM} & \mbox{fICA}\\
	 -0.50 & 0.02 & -0.03 & 0.11 & 0.52 & 0.43 & 0.54 & 0.52 & 0.43 & 0.55 \\
     0.87 & -0.23 & -0.16 & -0.19 & 0.32 & 0.36 & 0.25 & 0.40 & 0.42 & 0.30 \\
\bottomrule
\end{array}\]
\end{table}

\section{Application to SVAR}

\subsection{Structural Vector Autoregression}

Structural Vector Autoregressions (SVAR) have attracted a lot of interest in time series econometrics since the pioneering work of \citet{sims1980macroeconomics}. The standard model postulates that some observed state of the economy characterized by a vector of $n$ variables, $Y_t$, is related to unobserved (stationary) shocks (\textit{e.g.}, monetary or oil shocks) through

\begin{equation}
    Y_t = \Theta(L) \varepsilon_t
\end{equation}

where $\Theta(L)$ is an unknown lag polynomial that represents the impulse response function\footnote{Similarly to ICA systems, shocks as well as their effects on the system are unobserved so that there is a scale indeterminacy: shocks can be arbitrarily re-scaled to get an observationally equivalent system in which the effect of shocks are inversely re-scaled. Hence a normalization is typically imposed, for instance the unit variance normalization (each shock has variance one) or the unit effect normalization ($\Theta_{jj} = 1 \ \forall j$) are popular.}. $\Theta(L)$ describes the transmission mechanism of shocks to the economy and a subset of its column typically constitutes parameters of interest. \medskip

The first step towards estimation of $\Theta(L)$ is usually to perform the vector autoregression $A(L) Y_t = \eta_t$ to recover estimates of the innovation vector, $\eta_t$. The fundamentalness assumption states that the span of the shocks and innovations are identical and thus $\eta_t = \Theta \varepsilon_t$ for some invertible matrix $\Theta$. It is well-known in the literature (see for instance \citet{forni2019structural}) that $\Theta$ corresponds to the first term in the lag polynomial $\Theta(L)$. \medskip

Once $\Theta$ is identified, the whole lag polynomial is recovered as $A(L)^{-1} \Theta$. As a result, the problem is reduced to the system $\eta_t = \Theta \varepsilon_t$. While standard SVAR only assumes that entries in $\varepsilon_t$ are uncorrelated, second moments -- $\Sigma_\eta = \Theta \Sigma_\varepsilon \Theta'$ -- bring too few equations to solve for $\Theta$. Various solutions have been proposed, among which short-run restrictions \citep{sims1980macroeconomics}, long-run restrictions \citep{king1987stochastic, blanchard1989dynamic, shapiro1988sources}, identification by heteroskedasticity \citep{rigobon2003identification, sentana2001identification, lewis2019identifying}, or sign restrictions \citep{uhlig2005effects}. A good recent reference is \citet{kilian2017structural}. \medskip

Although these restrictions solve the identification problem, assuming a priori knowledge of numerous ($n(n-1)$) shocks' effects is often an issue, as the transmission mechanism of shocks to the economy is primarily an empirical question. Thus, some authors \citep{siegfried2002information, gourieroux2014revisiting} have pointed out that $\eta_t = \Theta \varepsilon_t$ can be identified by assuming that $\varepsilon_t$ contains non-Gaussian independent variables, bypassing restrictions on $\Theta$ and introducing ICA methods to the SVAR literature. Many subsequent studies have followed that road, estimating the model using high-order moments \citep{guay2018identification, keweloh2019generalized}, or pseudo-maximum of likelihood \citep{gourieroux2017statistical}. See also related discussions and methods in \citet{moneta2013causal, herwartz2015structural, lanne2017identification}. \medskip

\subsection{Application}

I consider a standard SVAR system with monthly data $Y_t$ on real economic, oil price, and stock market growths\footnote{Data comes from the following sources: \\ https://fred.stlouisfed.org/series/INDPRO (industrial production); \\ https://www.eia.gov/dnav/pet/hist/LeafHandler.ashx?n=pet\&s=r0000____3\&f=m (oil); \\ https://finance.yahoo.com/quote/\%5EGSPC?p=\%5EGSPC (SP);\\ https://fred.stlouisfed.org/series/CPIAUCSL (CPI)} as in \citet{keweloh2019generalized}. The first-step vector autoregression $A(L) Y_t = \eta_t$ is performed with four lags, as suggested by Akaike's information criterion. \medskip

The study is interesting to replicate for two reasons. First, as in many SVAR studies, there might be concerns about the fundamentalness assumption. This could for instance be caused by the presence of additional shocks (\textit{e.g.}, due to measurement error). Though robustness results against non-fundamentalness exist \citep{sims2006does, sims2012news, feve2012identifying, beaudry2015nonfundamentalness, forni2019structural}, it is worthwhile to see if the test detects a problem about the validity of the ICA representation. \medskip 

Second, shocks might have quite fat tails in practice. For instance, \citet{keweloh2019generalized} obtains excess kurtosis for all shocks and find that the shock associated to economic activity has a kurtosis above 10. Thus an estimator robust to existence of moment and able to perform accurate estimation in presence of fat tails may be useful. \medskip

The object of interest is here the lag polynomial $\Theta(L)$, rather than solely the unmixing matrix. I report the estimated responses to shock in figure 1 and display bootstraped confidence intervals. \medskip 

Shocks are here subject to the unit norm normalization, so they have the same overall variance over the system. Shocks 2 and 3 have similar variance of about 89, and affect strongly economic activity. The first shock accounts for less of the disturbances to the economic system (variance of 31) and has a lower contemporaneous effect on economic activity; it seems to affect the whole system negatively after a period, but the impact is imprecisely estimated. \medskip

The over-identification test' does not reject the null (p-value 0.21), so that there is no evidence against the validity of the ICA representation.

% Interesting to replicate this because (i) possible non-fundamentalness. Although there are robustness results, additional shocks could hinder ICA methods so worth testing. (ii) my method deals with fat tails, so might be worse a robustness check.  

\begin{figure}
\begin{subfigure}{.33\textwidth}
  \centering
  \includegraphics[width=.5\linewidth]{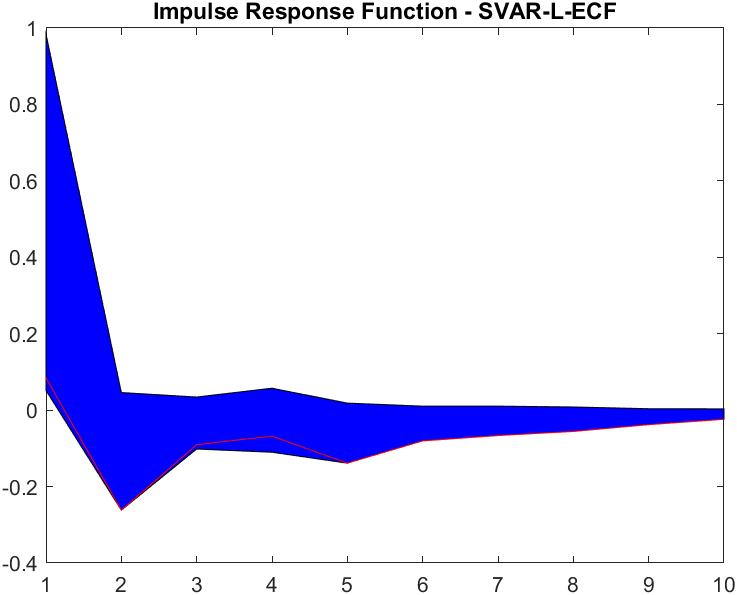}
  \label{fig:sfig1}
\end{subfigure}%
\begin{subfigure}{.33\textwidth}
  \centering
  \includegraphics[width=.5\linewidth]{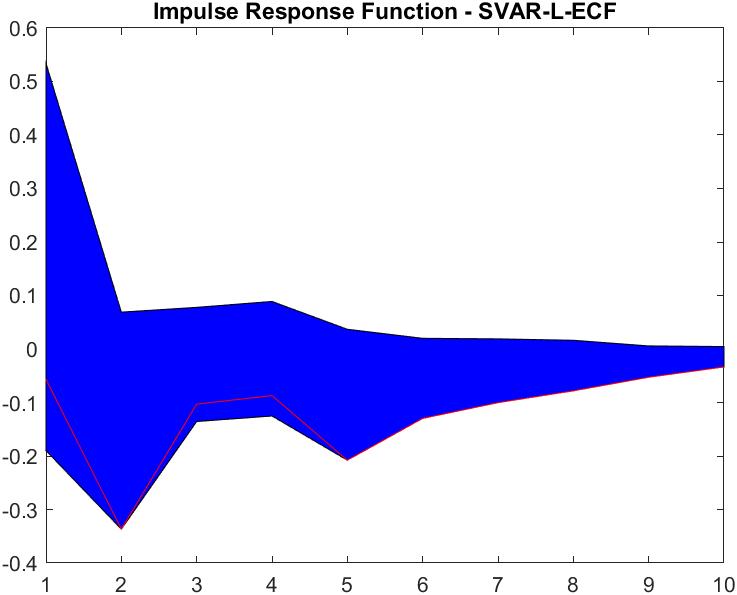}
  \label{fig:sfig2}
\end{subfigure}%
\begin{subfigure}{.33\textwidth}
  \centering
  \includegraphics[width=.5\linewidth]{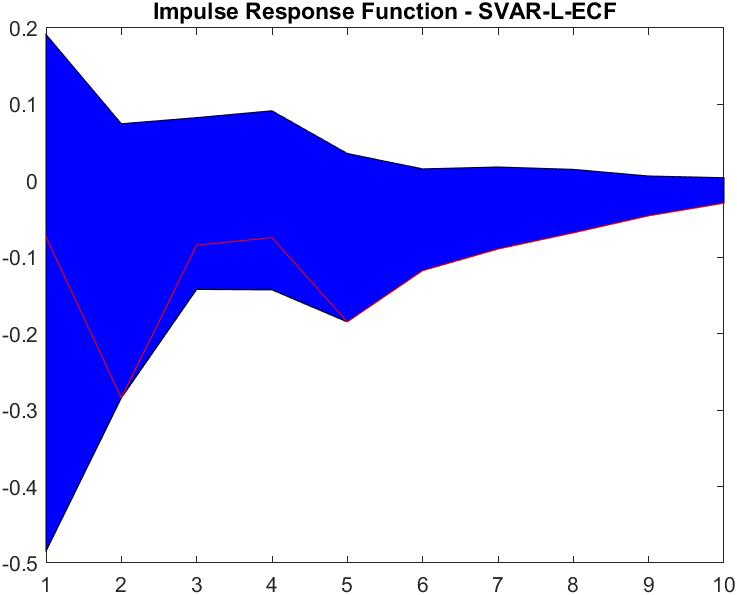}
  \label{fig:sfig3}
\end{subfigure} \\
\begin{subfigure}{.33\textwidth}
  \centering
  \includegraphics[width=.5\linewidth]{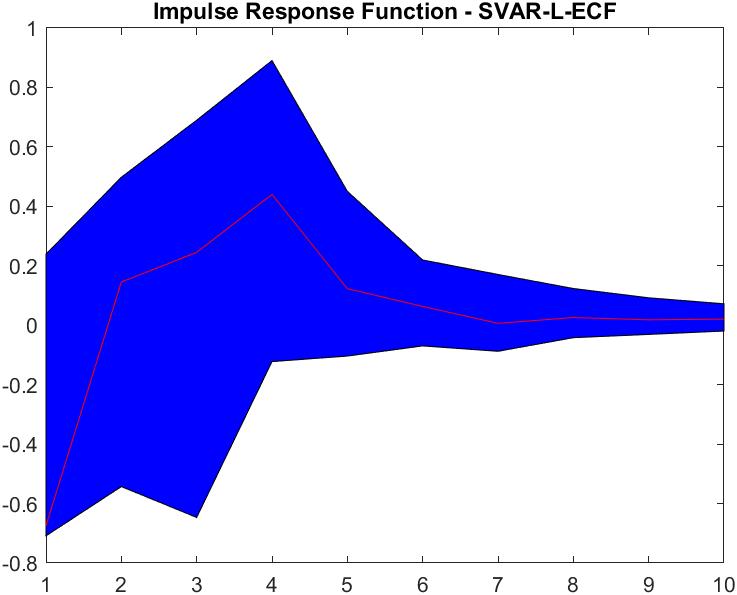}
  \label{fig:sfig4}
\end{subfigure}%
\begin{subfigure}{.33\textwidth}
  \centering
  \includegraphics[width=.5\linewidth]{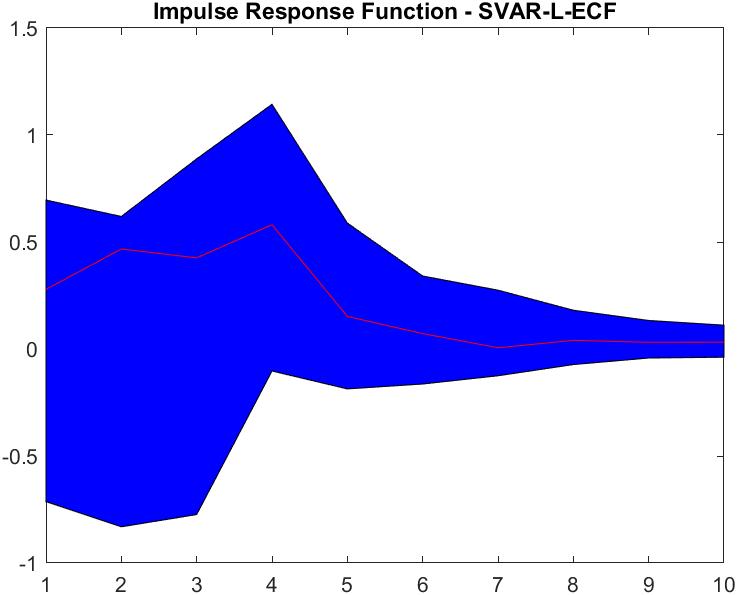}
  \label{fig:sfig5}
\end{subfigure}%
\begin{subfigure}{.33\textwidth}
  \centering
  \includegraphics[width=.5\linewidth]{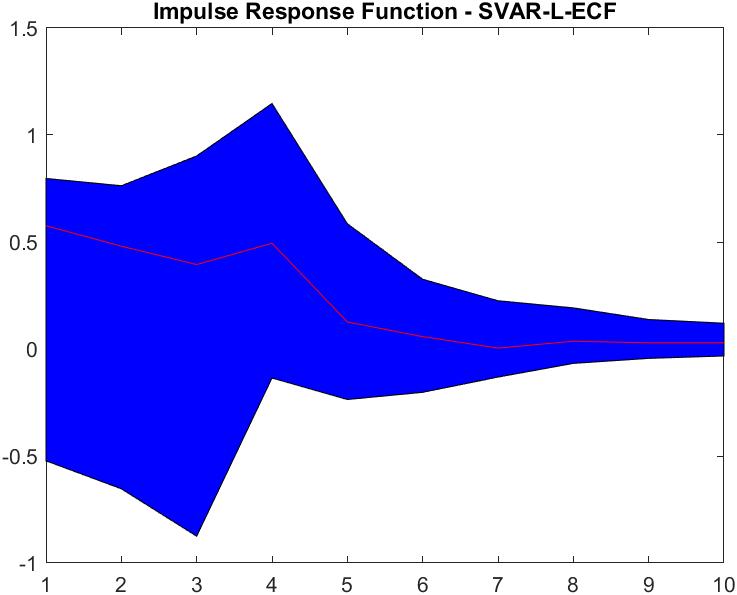}
  \label{fig:sfig6}
\end{subfigure} \\
\begin{subfigure}{.33\textwidth}
  \centering
  \includegraphics[width=.5\linewidth]{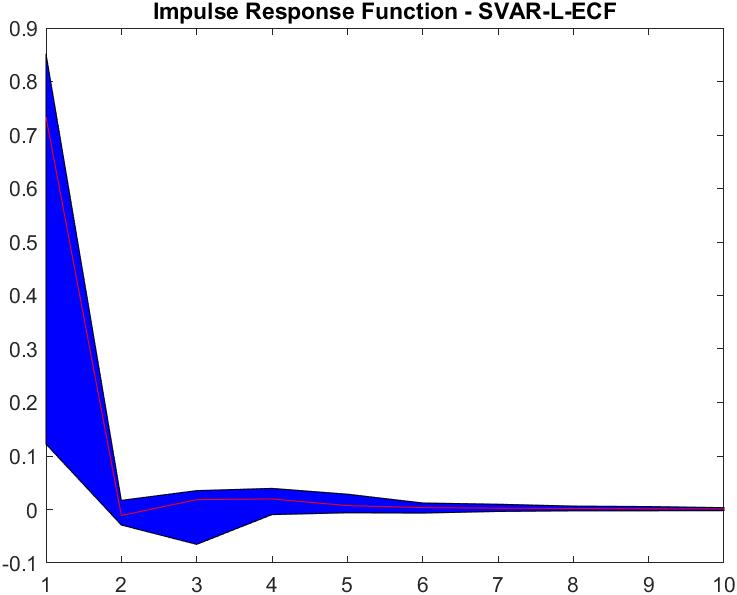}
  \label{fig:sfig7}
\end{subfigure}%
\begin{subfigure}{.33\textwidth}
  \centering
  \includegraphics[width=.5\linewidth]{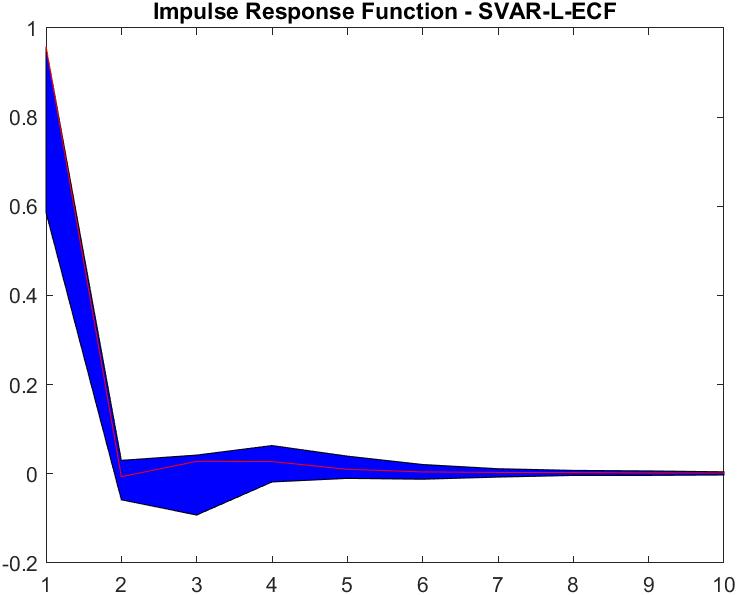}
  \label{fig:sfig8}
\end{subfigure}%
\begin{subfigure}{.33\textwidth}
  \centering
  \includegraphics[width=.5\linewidth]{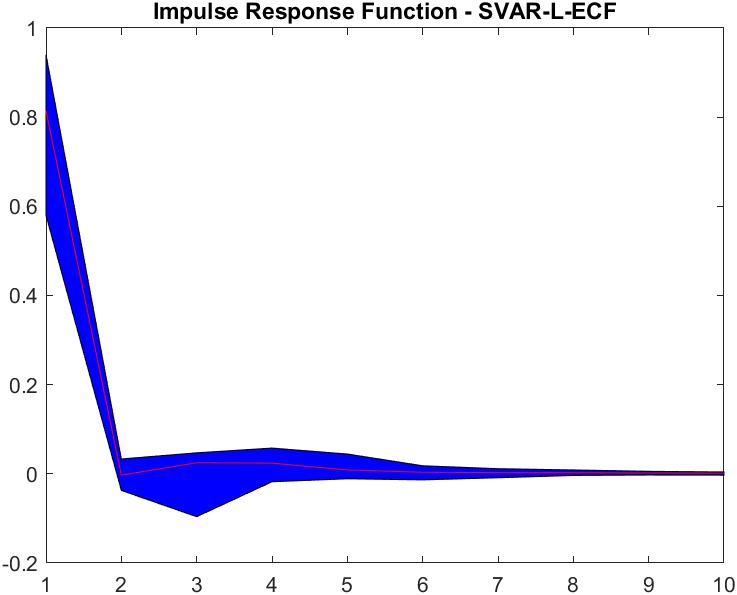}
  \label{fig:sfig9}
\end{subfigure}
\caption{Plots of Impulse Responses Functions. Each column represents the 1-to-10-months impact of a shock on the S\&P (first row), oil price (second row), and economic activity (third row). Shaded area depicts 90\% bootstrap confidence interval.}
\label{fig:fig}
\end{figure}

\pagebreak
\bibliographystyle{aea}
\bibliography{References}

@article{reiersol1950identifiability,
  title={Identifiability of a linear relation between variables which are subject to error},
  author={Reiers{\o}l, Olav},
  journal={Econometrica: Journal of the Econometric Society},
  pages={375--389},
  year={1950},
  publisher={JSTOR}
}

@article{keweloh2019generalized,
  title={A Generalized Method of Moments Estimator for Structural Vector Autoregressions Based on Higher Moments},
  author={Keweloh, Sascha Alexander},
  journal={Journal of Business \& Economic Statistics},
  pages={1--11},
  year={2019},
  publisher={Taylor \& Francis}
}

@article{comon1994independent,
  title={Independent component analysis, a new concept?},
  author={Comon, Pierre},
  journal={Signal processing},
  volume={36},
  number={3},
  pages={287--314},
  year={1994},
  publisher={Elsevier}
}

@article{eriksson2003identifiability,
  title={Identifiability and separability of linear ICA models revisited},
  author={Eriksson, Jan and Koivunen, Visa},
  journal={Proc. of ICA},
  volume={2003},
  pages={23--27},
  year={2003},
  publisher={none}
}

@article{darmois1953analyse,
  title={Analyse g{\'e}n{\'e}rale des liaisons stochastiques: étude particuli{\`e}re de l'analyse factorielle lin{\'e}aire},
  author={Darmois, George},
  journal={Revue de l'Institut international de statistique},
  pages={2--8},
  year={1953},
  publisher={JSTOR}
}

@article{sims1980macroeconomics,
  title={Macroeconomics and reality},
  author={Sims, Christopher A},
  journal={Econometrica: journal of the Econometric Society},
  pages={1--48},
  year={1980},
  publisher={JSTOR}
}

@article{newey1994large,
  title={Large sample estimation and hypothesis},
  author={Newey, KW and McFadden, D},
  journal={Handbook of Econometrics, IV, Edited by RF Engle and DL McFadden},
  pages={2112--2245},
  year={1994},
  publisher={Citeseer}
}

@article{feuerverger1977empirical,
  title={The empirical characteristic function and its applications},
  author={Feuerverger, Andrey and Mureika, Roman A and others},
  journal={The annals of Statistics},
  volume={5},
  number={1},
  pages={88--97},
  year={1977},
  publisher={Institute of Mathematical Statistics}
}

@article{csorgo1981limit,
  title={Limit behaviour of the empirical characteristic function},
  author={Csorgo, Sandor},
  journal={The Annals of Probability},
  pages={130--144},
  year={1981},
  publisher={JSTOR}
}

@techreport{ben2016identification,
  title={Identification and estimation of coefficients in dependent factor models},
  author={Ben-Moshe, Dan},
  year={2016},
  institution={Working paper, The Hebrew University of Jerusalem}
}

@article{blanchard1989dynamic,
  title={The dynamic effects of aggregate demand and aggregate supply},
  author={Blanchard, Olivier J and Quah, Danny},
  journal={The American Economic Review},
  volume={79},
  number={4},
  pages={655--673},
  year={1989}
}

@article{uhlig2005effects,
  title={What are the effects of monetary policy on output? Results from an agnostic identification procedure},
  author={Uhlig, Harald},
  journal={Journal of Monetary Economics},
  volume={52},
  number={2},
  pages={381--419},
  year={2005},
  publisher={Elsevier}
}

@article{shapiro1988sources,
  title={Sources of business cycle fluctuations},
  author={Shapiro, Matthew D and Watson, Mark W},
  journal={NBER Macroeconomics annual},
  volume={3},
  pages={111--148},
  year={1988},
  publisher={MIT Press}
}

@techreport{king1987stochastic,
  title={Stochastic trends and economic fluctuations},
  author={King, Robert G and Plosser, Charles I and Stock, James H and Watson, Mark W},
  year={1987},
  institution={National Bureau of Economic Research}
}

@article{forni2019structural,
  title={Structural VARs and noninvertible macroeconomic models},
  author={Forni, Mario and Gambetti, Luca and Sala, Luca},
  journal={Journal of Applied Econometrics},
  volume={34},
  number={2},
  pages={221--246},
  year={2019},
  publisher={Wiley Online Library}
}

@article{skitovitch1953property,
  title={On a property of the normal distribution},
  author={Skitovitch, Viktor P},
  journal={DAN SSSR},
  volume={89},
  pages={217--219},
  year={1953}
}

@article{feuerverger1990efficiency,
  title={An efficiency result for the empirical characteristic function in stationary time-series models},
  author={Feuerverger, Andrey},
  journal={The Canadian Journal of Statistics/La Revue Canadienne de Statistique},
  pages={155--161},
  year={1990},
  publisher={JSTOR}
}

@article{guay2018identification,
  title={Identification of structural vector autoregressions through higher unconditional moments},
  author={Guay, Alain},
  journal={Journal of Econometrics},
  volume={225},
  number={1},
  pages={27--46},
  year={2021},
  publisher={Elsevier}
}

@article{sims2006does,
  title={Does monetary policy generate recessions?},
  author={Sims, Christopher A and Zha, Tao},
  journal={Macroeconomic Dynamics},
  volume={10},
  number={2},
  pages={231--272},
  year={2006},
  publisher={Cambridge University Press}
}

@article{sims2012news,
  title={News, non-invertibility, and structural VARs},
  author={Sims, Eric R},
  journal={Advances in Econometrics},
  volume={28},
  pages={81},
  year={2012}
}

@article{feve2012identifying,
  title={Identifying news shocks from SVARs},
  author={Feve, Patrick and Jidoud, Ahmat},
  journal={Journal of Macroeconomics},
  volume={34},
  number={4},
  pages={919--932},
  year={2012},
  publisher={Elsevier}
}

@techreport{beaudry2015nonfundamentalness,
  title={When is nonfundamentalness in VARs a real problem? An application to news shocks},
  author={Beaudry, Paul and F{\`e}ve, Patrick and Guay, Alain and Portier, Franck},
  year={2015},
  institution={National Bureau of Economic Research}
}

@article{lewis2019identifying,
  title={Identifying shocks via time-varying volatility},
  author={Lewis, Daniel},
  journal={FRB of New York Staff Report},
  number={871},
  year={2019}
}

@article{rigobon2003identification,
  title={Identification through heteroskedasticity},
  author={Rigobon, Roberto},
  journal={Review of Economics and Statistics},
  volume={85},
  number={4},
  pages={777--792},
  year={2003},
  publisher={MIT Press}
}

@article{sentana2001identification,
  title={Identification, estimation and testing of conditionally heteroskedastic factor models},
  author={Sentana, Enrique and Fiorentini, Gabriele},
  journal={Journal of econometrics},
  volume={102},
  number={2},
  pages={143--164},
  year={2001},
  publisher={Elsevier}
}

@article{chen2005consistent,
  title={Consistent independent component analysis and prewhitening},
  author={Chen, Aiyou and Bickel, Peter J},
  journal={IEEE Transactions on Signal Processing},
  volume={53},
  number={10},
  pages={3625--3632},
  year={2005},
  publisher={IEEE}
}

@article{eriksson2003characteristic,
  title={Characteristic-function-based independent component analysis},
  author={Eriksson, Jan and Koivunen, Visa},
  journal={Signal Processing},
  volume={83},
  number={10},
  pages={2195--2208},
  year={2003},
  publisher={Elsevier}
}

@article{gourieroux2017statistical,
  title={Statistical inference for independent component analysis: Application to structural VAR models},
  author={Gouri{\'e}roux, Christian and Monfort, Alain and Renne, Jean-Paul},
  journal={Journal of Econometrics},
  volume={196},
  number={1},
  pages={111--126},
  year={2017},
  publisher={Elsevier}
}

@article{carrasco2017efficient,
  title={Efficient estimation using the characteristic function},
  author={Carrasco, Marine and Kotchoni, Rachidi},
  journal={Econometric Theory},
  volume={33},
  number={2},
  pages={479--526},
  year={2017},
  publisher={Cambridge University Press}
}

@article{carrasco2000generalization,
  title={Generalization of GMM to a continuum of moment conditions},
  author={Carrasco, Marine and Florens, Jean-Pierre},
  journal={Econometric Theory},
  pages={797--834},
  year={2000},
  publisher={JSTOR}
}

@article{bach2002kernel,
  title={Kernel independent component analysis},
  author={Bach, Francis R and Jordan, Michael I},
  journal={Journal of machine learning research},
  volume={3},
  number={Jul},
  pages={1--48},
  year={2002}
}

@article{chen2006efficient,
  title={Efficient independent component analysis},
  author={Chen, Aiyou and Bickel, Peter J and others},
  journal={The Annals of Statistics},
  volume={34},
  number={6},
  pages={2825--2855},
  year={2006},
  publisher={Institute of Mathematical Statistics}
}

@article{ilmonen2011semiparametrically,
  title={Semiparametrically efficient inference based on signed ranks in symmetric independent component models},
  author={Ilmonen, Pauliina and Paindaveine, Davy and others},
  journal={the Annals of Statistics},
  volume={39},
  number={5},
  pages={2448--2476},
  year={2011},
  publisher={Institute of Mathematical Statistics}
}

@article{samworth2012independent,
  title={Independent component analysis via nonparametric maximum likelihood estimation},
  author={Samworth, Richard J and Yuan, Ming and others},
  journal={The Annals of Statistics},
  volume={40},
  number={6},
  pages={2973--3002},
  year={2012},
  publisher={Institute of Mathematical Statistics}
}

@article{oja2006fastica,
  title={The FastICA algorithm revisited: Convergence analysis},
  author={Oja, Erkki and Yuan, Zhijian},
  journal={IEEE Transactions on Neural Networks},
  volume={17},
  number={6},
  pages={1370--1381},
  year={2006},
  publisher={IEEE}
}

@book{kilian2017structural,
  title={Structural vector autoregressive analysis},
  author={Kilian, Lutz and L{\"u}tkepohl, Helmut},
  year={2017},
  publisher={Cambridge University Press}
}

@article{hyvarinen2001independent,
  title={al. Independent component analysis. John Willey ans Sons},
  author={Hyv{\"a}rinen, Aapo and Karhunen, Juha and Oja, Erkki},
  journal={Inc., New York},
  year={2001}
}

@article{amengual2020testing,
  title={Testing distributional assumptions using a continuum of moments},
  author={Amengual, Dante and Carrasco, Marine and Sentana, Enrique},
  journal={Journal of Econometrics},
  volume={218},
  number={2},
  pages={655--689},
  year={2020},
  publisher={Elsevier}
}

@article{samarov2004nonparametric,
  title={Nonparametric independent component analysis},
  author={Samarov, Alexander and Tsybakov, Alexandre and others},
  journal={Bernoulli},
  volume={10},
  number={4},
  pages={565--582},
  year={2004},
  publisher={Bernoulli Society for Mathematical Statistics and Probability}
}

@techreport{gourieroux2014revisiting,
  title={Revisiting identification and estimation in structural VARMA models},
  author={Gourieroux, Christian and Monfort, Alain and others},
  year={2014}
}

@article{lanne2017identification,
  title={Identification and estimation of non-Gaussian structural vector autoregressions},
  author={Lanne, Markku and Meitz, Mika and Saikkonen, Pentti},
  journal={Journal of Econometrics},
  volume={196},
  number={2},
  pages={288--304},
  year={2017},
  publisher={Elsevier}
}

@article{moneta2013causal,
  title={Causal inference by independent component analysis: Theory and applications},
  author={Moneta, Alessio and Entner, Doris and Hoyer, Patrik O and Coad, Alex},
  journal={Oxford Bulletin of Economics and Statistics},
  volume={75},
  number={5},
  pages={705--730},
  year={2013},
  publisher={Wiley Online Library}
}

@techreport{siegfried2002information,
  title={An information-theoretic extension to structural VAR modelling},
  author={Siegfried, Nikolaus A and others},
  year={2002},
  institution={Hamburg University, Department of Economics}
}

@article{herwartz2015structural,
  title={Structural VAR Modelling with Independent Innovations—An Analysis of Macroeconomic Dynamics in the Euro Area Based on a Novel Identification Scheme},
  author={Herwartz, H},
  journal={University of G{\"o}ttingen, Working Paper},
  year={2015}
}

@article{ablin2018faster,
  title={Faster independent component analysis by preconditioning with Hessian approximations},
  author={Ablin, Pierre and Cardoso, Jean-Fran{\c{c}}ois and Gramfort, Alexandre},
  journal={IEEE Transactions on Signal Processing},
  volume={66},
  number={15},
  pages={4040--4049},
  year={2018},
  publisher={IEEE}
}

@article{bonhomme2009consistent,
  title={Consistent noisy independent component analysis},
  author={Bonhomme, St{\'e}phane and Robin, Jean-Marc},
  journal={Journal of Econometrics},
  volume={149},
  number={1},
  pages={12--25},
  year={2009},
  publisher={Elsevier}
}

@article{hansen1982large,
  title={Large sample properties of generalized method of moments estimators},
  author={Hansen, Lars Peter},
  journal={Econometrica: Journal of the Econometric Society},
  pages={1029--1054},
  year={1982},
  publisher={JSTOR}
}

@article{schennach2004estimation,
  title={Estimation of nonlinear models with measurement error},
  author={Schennach, Susanne M},
  journal={Econometrica},
  volume={72},
  number={1},
  pages={33--75},
  year={2004},
  publisher={Wiley Online Library}
}

@article{carrasco2002efficient,
  title={Efficient GMM estimation using the empirical characteristic function},
  author={Carrasco, Marine and Florens, Jean-Pierre},
  year={2002},
  publisher={IDEI Working paper}
}

@inproceedings{agrawal2022blind,
  title={Blind source separation in perspective of ica algorithms: A review},
  author={Agrawal, Jharna and Gupta, Manish and Garg, Hitendra},
  booktitle={2022 International Conference on Computational Intelligence and Sustainable Engineering Solutions (CISES)},
  pages={78--85},
  year={2022},
  organization={IEEE}
}

\section{Appendix A: proofs}

\subsection{Theorem 3.1 (Consistency)}

\begin{proof}

Consistency follows by verifying assumptions of Theorem 2.1 in \citet{newey1994large}. The parameter space is compact by construction, identification is established, and the limiting objective function is continuous by dominated convergence since characteristic functions are bounded. It remains to show uniform convergence in probability. \medskip

The empirical characteristic function using a consistent estimator of $\eta_t$ converges uniformly in probability: 

\begin{align*}
\begin{split}
    \sup_{\Theta \in \overline{\Theta}} \left|\frac{1}{T} \sum_{t=1}^T e^{i \vec{s} P_j \hat{\eta}_t}  - \mathds{E}[e^{i \vec{s} P_j \eta_t}] \right| & = \sup_{\Theta \in \overline{\Theta}} \left|\frac{1}{T} \sum_{t=1}^T e^{i \vec{s} P_j \hat{\eta}_t} - \frac{1}{T} \sum_{t=1}^T e^{i \vec{s} P_j \eta_t} + \frac{1}{T} \sum_{t=1}^T e^{i \vec{s} P_j \eta_t} - \mathds{E}[e^{i \vec{s} P_j \eta_t}] \right| \\ &
    \leq \sup_{\Theta \in \overline{\Theta}} \left|\frac{1}{T} \sum_{t=1}^T e^{i \vec{s} P_j \hat{\eta}_t} - \frac{1}{T} \sum_{t=1}^T e^{i \vec{s} P_j \eta_t} \right| + \sup_{\Theta \in \overline{\Theta}} \left|\frac{1}{T} \sum_{t=1}^T e^{i \vec{s} P_j \eta_t} - \mathds{E}[e^{i \vec{s} P_j \eta_t}] \right| \\ &
    \leq \sup_{\Theta \in \overline{\Theta}} \left|\frac{1}{T} \sum_{t=1}^T e^{i \vec{s} P_j \eta_t} (e^{i \vec{s} P_j (\hat{\eta}_t-\eta_t)} - 1) \right| + o_p(1) \\ &
    \leq \sup_{\Theta \in \overline{\Theta}} \frac{1}{T} \sum_{t=1}^T \left|e^{i \vec{s} P_j (\hat{\eta}_t-\eta_t)} - 1 \right| + o_p(1) \\ &
    \leq \sup_{\Theta \in \overline{\Theta}} \frac{1}{T} \sum_{t=1}^T \left|\vec{s} P_j (\hat{\eta}_t-\eta_t) \right| + o_p(1) \\ &
    \leq \frac{1}{T} \sum_{t=1}^T \Vert \vec{s} \Vert \frac{1}{\cos(\frac{\pi}{2}-\epsilon)} |\hat{\eta}_t-\eta_t| + o_p(1) \\ & 
    \rightarrow^p 0
\end{split}
\end{align*}

where the second inequality follows from the uniform law of large numbers, the fourth from $|e^{ix}-1| \leq |x|$, the fifth from $\epsilon$-invertibility, and the convergence is implied by consistency of $\hat{\eta}_t$. \medskip

Hence, by Theorem 2.1 in \citet{newey1994large}, $\hat{\theta} \rightarrow^p \theta_0$. \end{proof}

\subsection{Theorem 3.2 (Asymptotic Normality)}

\begin{proof}

Derivation of asymptotic normality follows the approach in \citet{newey1994large}. The arguments of integration $\vec{s}$, $\vec{r}$ do not play a significant role and are suppressed to ease notation. By dominated convergence, first-order conditions read
\begin{equation}
    \int  \begin{pmatrix} \Re\frac{\partial q_T}{\partial \theta} (\hat{\theta}) & \Im(\frac{\partial q_T}{\partial \theta} (\hat{\theta})) \end{pmatrix} W_T \begin{pmatrix} \Re q_T(\hat{\theta}) \\ \Im q_T(\hat{\theta})\end{pmatrix} = 0
\end{equation}

Applying the mean-value theorem around the true value $\theta_0$ yields
\begin{align*}
\begin{split}
    \int  \begin{pmatrix} \Re\frac{\partial q_T}{\partial \theta} (\hat{\theta}) & \Im(\frac{\partial q_T}{\partial \theta} (\hat{\theta})) \end{pmatrix} W_T \begin{pmatrix} \Re q_T(\theta_0) + (\Re \frac{\partial q_T}{\partial \theta} (\tilde{\theta}_R))^' (\hat{\theta} - \theta_0) \\ \Im q_T(\theta_0) + (\Im \frac{\partial q_T}{\partial \theta} (\tilde{\theta}_I))^' (\hat{\theta} - \theta_0) \end{pmatrix} = 0
\end{split}
\end{align*}

so that, rearranging
\begin{align*}
\begin{split}
    \sqrt{T} (\hat{\theta} - \theta_0) =
    - \left[\int  \begin{pmatrix} \Re\frac{\partial q_T}{\partial \theta} (\hat{\theta}) & \Im(\frac{\partial q_T}{\partial \theta} (\hat{\theta})) \end{pmatrix} W_T \begin{pmatrix} (\Re \frac{\partial q_T}{\partial \theta} (\tilde{\theta}_R))^' \\ (\Im \frac{\partial q_T}{\partial \theta} (\tilde{\theta}_I))^' \end{pmatrix} \right]^{-1} \\ 
    \sqrt{T} \int  \begin{pmatrix} \Re\frac{\partial q_T}{\partial \theta} (\hat{\theta}) & \Im(\frac{\partial q_T}{\partial \theta} (\hat{\theta})) \end{pmatrix} W_T \begin{pmatrix} \Re q_T(\theta_0) \\ \Im q_T(\theta_0)\end{pmatrix} 
\end{split}
\end{align*}

The first term can be proven to converge uniformly in probability, proceeding as in the uniform convergence step in proving consistency. For the second term, the empirical characteristic function converges to a complex normal stochastic process. This follows from convergence of finite dimensional distributions (by the multivariate central limit theorem and the delta method) and tightness (see \citet{feuerverger1977empirical}\footnote{As pointed out by \citet{csorgo1981limit}, the result requires slightly stronger conditions than initially thought. Existence of moments larger than 1 suffices.}\footnote{They also proved almost sure convergence of $\sup_{-K \leq s \leq K} |c_n(s) - c(s)|$, where $c$ denotes the empirical characteristic function and $c_n$ its empirical counterpart, the empirical characteristic function is almost surely bounded away from zero on a neighborhood of the origin for $T$ large enough, and tightness can thus be established for the cumulant generating function as well. As a result, the log-version of the criterion can also be shown to be asymptotically normal}). \medskip

Then convergence of $\sqrt{T} (\frac{1}{T} \sum_{i=1}^T e^{i \vec{v} \eta_t} - \mathds{E}[e^{i \vec{v} \eta_t}])$ to a complex normal stochastic process together with the continuous mapping theorem deliver asymptotic normality. If $\eta_t$ is known, we directly obtain
\begin{equation*}
    \sqrt{T} (\hat{\theta} - \theta_0) \overset{d}{\rightarrow} N(0; B V B')
\end{equation*}

Interestingly, estimation of $\eta_t$ does not affect the asymptotic variance. Indeed, estimation of $\eta_t$ can be accounted for by expanding $q_0(\hat{\eta}_t)=q_0(\eta_t(\hat{\beta}))$ into $q_0(\eta_t(\beta_0)) + \frac{\partial q(\theta_0, \eta_t(\beta_0))}{\partial \beta} (\hat{\beta} - \beta)$ plus lower-order terms, where $\beta$ is the underlying parameter vector in estimating $\eta_t$. The first term corresponds to the case where $\eta_t$ is observed. If $\eta_t$ is an error term independent from its regressors $w_t$, then by properties of the $P_j$, 
\begin{align*}
    \frac{\partial q_0}{\partial \beta} & = \varphi(\vec{s}) \sum_{j=0}^n a_j \frac{\mathds{E}[e^{i \vec{s} P_j \eta_t} (-w_t) P_j' \vec{s}']}{\mathds{E}[e^{i \vec{s} P_j \eta_t}]} \\
    & = - \varphi(\vec{s}) \mathds{E}[w_t] \left(\sum_{j=0}^n a_j P_j'\right) \vec{s}' \\
    & = 0
\end{align*}

\end{proof}

\pagebreak

\section*{Appendix B: Additional results}

\subsection*{The asymptotic variance}

Due to the asymptotic linear representation shown while establishing asymptotic normality, the variance can be approximated via bootstrap. Alternatively, the asymptotic variance can be consistently estimated. Indeed, most terms appearing in its expression have a natural estimator based on the use of $\hat{\theta}$ in place of $\theta_0$ and the use of the sample counterparts of population quantities. 

Assume for ease of exposition that the characteristic functions do not vanish. This simplifies slightly the following formulae by allowing characteristic functions in the denominators, but is not necessary. We have

\begin{equation}
    \frac{\partial q_0}{\partial \theta^{'}} = - \varphi(\vec{s}) \sum_{j=1}^n  \frac{\partial \ln(\mathds{E}[e^{i \vec{s} P_j \eta_t}])}{\partial \theta^{'}} = - \varphi(\vec{s}) \sum_{j=1}^n \frac{\mathds{E}\left[e^{i \vec{s} P_j \eta_t} (\eta_t^{'} \otimes \vec{s}) \frac{\partial \vect(P_j)}{\partial \theta^{'}}\right]}{\mathds{E}[e^{i \vec{s} P_j \eta_t}]}
\end{equation}

As an illustration of the differentiated term, consider the two-dimensional case:
\begin{equation}
    \Theta = \begin{pmatrix} \cos(\theta_1) & \cos(\theta_2) \\ \sin(\theta_1) & \sin(\theta_2) \end{pmatrix}
\end{equation}

Tedious but straightforward algebra yields 

\begin{equation}
    \Theta^{-1} = \frac{1}{\sin(\theta_2 - \theta_1)} \begin{pmatrix} \sin(\theta_2) & -\cos(\theta_2) \\ -\sin(\theta_1) & \cos(\theta_1) \end{pmatrix}
\end{equation}
\begin{equation}
    P_1 = \frac{1}{\sin(\theta_2 - \theta_1)}       \begin{pmatrix} \cos(\theta_1) \sin(\theta_2) & - \cos(\theta_1) \cos(\theta_2) \\ \sin(\theta_1) \sin(\theta_2) & -\sin(\theta_1) \cos(\theta_2) \end{pmatrix}
\end{equation}
\begin{equation}
    P_2 = \frac{1}{\sin(\theta_2 - \theta_1)}       \begin{pmatrix} - \sin(\theta_1) \cos(\theta_2) & \cos(\theta_1) \cos(\theta_2) \\ - \sin(\theta_1) \sin(\theta_2) & \cos(\theta_1) \sin(\theta_2) \end{pmatrix}
\end{equation}
\begin{equation}
    \frac{\partial \vect(P_1)}{\partial \theta^{'}} = \frac{1}{\sin^2(\theta_2 - \theta_1)} \begin{pmatrix} \sin(\theta_2) \cos(\theta_2) & - \sin(\theta_1) \cos(\theta_1) \\ \sin^2(\theta_2) & - \sin^2(\theta_1) \\ -\cos^2(\theta_2) & \cos^2(\theta_1) \\ -\sin(\theta_2) \cos(\theta_2) & \sin(\theta_1) \cos(\theta_1) \end{pmatrix}
\end{equation}
\begin{equation}
    \frac{\partial \vect(P_2)}{\partial \theta^{'}} = \frac{\partial \vect(I - P_1)}{\partial \theta^{'}} = - \frac{\partial \vect(P_1)}{\partial \theta^{'}}
\end{equation}

Hence, replacing expectations by sample averages and using consistent estimators in place of unknown parameters delivers a consistent estimate of $\frac{\partial q_0}{\partial \theta^{'}}$. \medskip 

Regarding the central term, the (centered) empirical characteristic function converges to a mean zero process with covariance function 
$\varphi(\vec{s} + \vec{r}) - \varphi(\vec{s}) \varphi(\vec{r})$. The covariance functions is enough to characterize the complex process noting that $\overline{\varphi(\vec{s})} = \varphi(-\vec{s})$. \medskip

Letting $\varPhi(\vec{s}) \equaldef \begin{pmatrix} \varphi(\vec{s}) \\ \varphi(\vec{s}_1) \\ ... \\ \varphi(\vec{s}_n) \end{pmatrix}$, where $\vec{s}_j \equaldef \vec{s} P_j$, and $A \equaldef \begin{pmatrix} a_0 & a_1 & ... & a_n \end{pmatrix}$, the covariances are obtained from 
\begin{align*}
\begin{split}
& \mathds{C}ov \left(\begin{pmatrix}\Re q_0(\vec{u}) \\ \Im q(\theta_0, \vec{u})\end{pmatrix}; \begin{pmatrix}\Re q_0(\vec{v}) \\ \Im q(\theta_0, , \vec{v})\end{pmatrix} \right) \\
& = \begin{pmatrix} A & 0 \\ 0 & A \end{pmatrix} \mathds{C}ov \left(\begin{pmatrix}\Re \varPhi(\vec{u}) \\ \Im \varPhi(\vec{u}) \end{pmatrix}; \begin{pmatrix}\Re \varPhi(\vec{v}) \\ \Im \varPhi(\vec{v}) \end{pmatrix} \right) \begin{pmatrix} A' & 0 \\ 0 & A' \end{pmatrix}
\end{split}
\end{align*}
where, using properties of the complex-normal distribution,
\begin{equation}
    \mathds{C}ov(\Re \varphi(\vec{u}_j); \Re \varphi(\vec{v}_k)) = \frac{1}{2} \Re \left(\frac{\varphi(\vec{u}_j-\vec{v}_k)}{\varphi(\vec{u}_j) \varphi(- \vec{v}_k)} + \frac{\varphi(\vec{u}_j+\vec{v}_k)}{\varphi(\vec{u}_j) \varphi(\vec{v}_k)} - 2 \right)
\end{equation}
\begin{equation}
    \mathds{C}ov(\Re \varphi(\vec{u}_j); \Im \varphi(\vec{v}_k)) = \frac{1}{2} \Im \left(- \frac{\varphi(\vec{u}_j-\vec{v}_k)}{\varphi(\vec{u}_j) \varphi(- \vec{v}_k)} + \frac{\varphi(\vec{u}_j+\vec{v}_k)}{\varphi(\vec{u}_j) \varphi(\vec{v}_k)} \right)
\end{equation}
\begin{equation}
    \mathds{C}ov(\Im \varphi(\vec{u}_j); \Re \varphi(\vec{v}_k)) = \frac{1}{2} \Im \left(\frac{\varphi(\vec{u}_j-\vec{v}_k)}{\varphi(\vec{u}_j) \varphi(- \vec{v}_k)} + \frac{\varphi(\vec{u}_j+\vec{v}_k)}{\varphi(\vec{u}_j) \varphi(\vec{v}_k)} - 2 \right)
\end{equation}
\begin{equation}
    \mathds{C}ov(\Im \varphi(\vec{u}_j); \Im \varphi(\vec{v}_k)) = \frac{1}{2} \Re \left(\frac{\varphi(\vec{u}_j-\vec{v}_k)}{\varphi(\vec{u}_j) \varphi(- \vec{v}_k)} - \frac{\varphi(\vec{u}_j+\vec{v}_k)}{\varphi(\vec{u}_j) \varphi(\vec{v}_k)} \right)
\end{equation}

All terms can be consistently estimated by using empirical characteristic functions as estimates of their population counterparts. \medskip

\textbf{Remark}: The covariance kernel is easier to handle directly in complex form. Let $\phi_{j}(\vec{s}) \equaldef \prod_{m \neq j} \hat{\varphi}(\vec{s} P_m)$ with $\phi_0(\vec{s}) = 1$. The covariance kernel is given by $K(\vec{r}, \vec{s}) = \sum_{k=0}^n \sum_{j=0}^n a_k a_j \phi_{j}(\vec{s}) \phi_{k}(-\vec{r}) (\varphi(\vec{r}_k-\vec{s}_j) - \varphi(\vec{r}_k) \varphi(\vec{s}_j))$. This is also valid when a characteristic function vanishes. \medskip

\textbf{Remark}: Similar results can be derived for the log version of the condition, \textit{i.e.}, $\sum_{j=0}^n a_j \ln\left(\frac{1}{T} \sum_{t=1}^T e^{i \vec{s} P_j \eta_t}\right) = 0$. The centered empirical log-characteristic function converges to a zero mean process with covariance function $\frac{\varphi(\vec{s} + \vec{r})}{\varphi(\vec{s}) \varphi(\vec{r})} - 1$ and satisfies $\overline{\ln(\varphi(\vec{s}))} = \ln(\varphi(- \vec{s}))$. The covariance kernel is then given by $\sum_{k=0}^n \sum_{j=0}^n a_k a_j \frac{\varphi(\vec{r}_k-\vec{s}_j)}{\varphi(\vec{r}_k) \varphi(- \vec{s}_j))}$.
\medskip

\end{document}